\documentclass[letter]{aa} 

\usepackage{graphicx}
\usepackage{txfonts}
\usepackage{amsmath}
\usepackage{bm}
\usepackage[breaklinks, colorlinks, citecolor=blue]{hyperref}
\newcommand{\healpix}{{\tt HEALPix}}
\newcommand{\Planck}{{\it Planck}}

\begin{document}

   \title{The multiphase and magnetized neutral hydrogen seen by LOFAR}

   \author{A. Bracco\inst{1}, 
            V. Jeli\'c\inst{1}, 
            A. Marchal\inst{2}, 
            L. Turi\'c\inst{1},
            A. Erceg\inst{1},
            M.-A. Miville-Desch$\hat{\rm e}$nes\inst{3}, \and 
            E. Bellomi\inst{4,5}
          }
  \institute{Ru{\dj}er Bo\v{s}kovi\'c Institute, Bijeni\v{c}ka cesta 54, 10 000 Zagreb, Croatia \\
                \email{abracco@irb.hr}
                \and
                Canadian Institute for Theoretical Astrophysics, University of Toronto, Toronto, ON M5S 3H8, Canada \and 
                Laboratoire AIM, CEA / CNRS / Universit\'e Paris-Saclay, 91191 Gif-sur-Yvette, France
                \and
 Laboratoire de Physique de l’Ecole Normale Supérieure, ENS, Université PSL, CNRS, Sorbonne Université, Université de Paris, Paris, France
         \and
 Observatoire de Paris, LERMA, 75005 Paris, France\\
}
\date{Received: 28 August 2020; Accepted: 10 November 2020}

  \abstract
  {Faraday tomography of polarimetric observations at low frequency in the radio is a unique tool for studying the structure of the magneto-ionic diffuse interstellar medium (ISM) based on Faraday depth. LOFAR data below 200 MHz have revealed a plethora of features in polarization, whose origin remains unknown. Previous studies have highlighted the remarkable association of such features with tracers of the magnetized-neutral ISM, such as interstellar dust and atomic hydrogen (HI). However, the physical conditions responsible for the correlation between magneto-ionic and neutral media have not been clarified yet. In this letter we further investigate the correlation between LOFAR data and the HI spectroscopic observations at 21cm from the {\it  Effelsberg-Bonn HI Survey} (EBHIS). We focus on the multiphase properties of the HI gas. We present the first statistical study on the morphological correlation between LOFAR tomographic data and the cold (CNM), lukewarm (LNM), and warm (WNM) neutral medium HI phases. We use the {\it Regularized Optimization for Hyper-Spectral Analysis} ({\tt ROHSA}) approach to decompose the HI phases based on a Gaussian decomposition of the HI spectra. We study four fields of view -- Fields 3C196, A, B, and C -- and find, in at least the first two, a significant correlation between the LOFAR and EBHIS data using the {\it histograms of oriented gradients} (HOG) feature. The absence of a correlation in Fields B and C is caused by a low signal-to-noise ratio in polarization. The observed HOG correlation in Fields 3C196 and A is associated with all HI phases and it is surprisingly dominant in the CNM and LNM phases.  We discuss possible mechanisms that would explain the correlation between CNM, LNM, and WNM with polarized emission at Faraday depths up to $10$ rad m$^{-2}$. Our results show how the complex structure of the ionic medium seen by the LOFAR tomographic data is tightly related to phase transition in the diffuse and magnetized neutral ISM traced by HI spectroscopic data.}


\authorrunning{A. Bracco et al.}
\titlerunning{Magnetized and diffuse neutral media seen by LOFAR}    
   \maketitle
%
\section{Introduction}\label{sec:intro}

Magnetic fields are among the most relevant, though relatively unknown, players of Galactic dynamics. They are key ingredients in the magnetohydrodynamic (MHD) turbulent cascade in the multiphase interstellar medium (ISM) that structures matter from the large kiloparsec scales in the Milky Way \citep{Brandenburg2005,Shukurov2006,Brandenburg2013,Beck2019,Ferriere2020} down to the parsec and subparsec scales of molecular clouds, where star formation in the Galaxy takes place \citep{Hennebelle2019}. 

Observations of synchrotron polarized and continuum emission at radio wavelengths \citep{Haslam1982,Reich1986,Davies1996}, and Faraday rotation represent powerful diagnostics for probing Galactic magnetic fields in the diffuse ISM \citep{Beck2015}. Clouds of thermal electrons in highly ionized and magnetized interstellar gas along the line of sight Faraday-rotate the Galactic synchrotron linear polarization. This highlights the structure of the turbulent magneto-ionic medium that features a network of "spaghetti-like" structures of polarized intensity and depolarization canals \citep[e.g.,][]{Haverkorn2003b, Haverkorn2003a, Haverkorn2003c,Gaensler2011,Iacobelli2014}.  

The full complexity of the magneto-ionic ISM has only been revealed, however, by Faraday tomography \citep{Burn1966,Brentjens2005}. This technique takes radio-polarimetric data and decomposes the observed polarized synchrotron emission by the amount of Faraday rotation it experiences along the line of sight. Faraday tomography maps the 3D relative distribution of the intervening magneto-ionic ISM based on Faraday depth (hereafter, $\phi$), or the cumulative effect of magnetic fields and thermal-electron density along the line of sight.  

Faraday tomography has been applied at different frequency ranges in the radio, from a few gigahertz \citep[i.e.,][]{Dickey2019} down to 400 MHz \citep{Thomson2019, Wolleben2019, Dickey2019} and below \citep{Jelic2014, Jelic2015, Lenc2016, vanEck2017, VanEck2019}. Because of the frequency dependence of Faraday rotation, the true power of Faraday tomography is seen at about $150$ MHz with the LOw Frequency ARray \citep[LOFAR,][]{vanHaarlem2013}, which is the instrument most sensitive to small column densities of magneto-ionic media, notably the warm ionized medium (WIM, completely ionized at temperatures of $\sim$10$^4$ K, \cite{Ferriere2020}).

LOFAR polarimetric observations are revealing a plethora of structures whose origins remain unknown. In Field 3C196,  \citet{Zaroubi2015} reported a striking correlation between LOFAR tomographic data and the orientation of the plane-of-the-sky Galactic magnetic field probed by polarized emission of interstellar dust \citep[i.e.,][]{Davis1951,Hall1949,PlanckIII2020}. This correlation became a puzzle, as dust total emission \citep{Boulanger1996} and polarization \citep{PIPXXXII, PIPXXXV2016, planck2015-XXXVIII,Soler2017b} are well known to be associated with cold gas in the Galaxy, mostly atomic and molecular, rather than with the WIM. A similar association, based on dust extinction only, was put forward in \citet{vanEck2017} based on the analysis of diffuse Galactic polarization from LOFAR toward the field centered on the galaxy IC 342.  

Spectroscopic emission of atomic hydrogen at 21 cm (HI) was also visually found correlated with the LOFAR polarized intensity in Field 3C196 \citep{Kalberla2016}, as well as with depolarization canals \citet{Jelic2018}. The HI medium is a mixture of bi-stable gas composed of a warm neutral medium (WNM), at temperatures of $\sim$8000 K and densities of $\sim$ 0.3 cm$^{-3}$, and a cold neutral medium (CNM), with corresponding temperatures and densities of $\sim$50 K and $\sim$ 50 cm$^{-3}$, respectively \citep{Field1965,Wolfire2003}. This atomic gas, subject to thermal instability, is also partially made up -- by up to 30\% -- of an unstable lukewarm neutral medium (LNM) with intermediate properties between the two stable phases \citep{Saury2014}. Magnetic fields have been strongly suggested to govern the formation of CNM structures, which are seen elongated as "fibers" along the magnetic-field orientation traced by dust polarization \citep{Clark2015,Clark2019}. The evolution of the multiphase and magnetized atomic medium toward the formation of CNM gas is likely at the origin of the observed correlation initially found by \citet{Zaroubi2015}. 
 
In this letter we further investigate the correlation between LOFAR data and the multiphase HI signal. We present the first statistical study that explicitly decomposes the latter into its CNM, LNM, and WNM components, and explores their respective association with the LOFAR polarized emission derived from Faraday tomography.  

\section{Description of the data}\label{sec:data}
\begin{figure*}[!ht]
\centering
\includegraphics[width=0.9\textwidth]{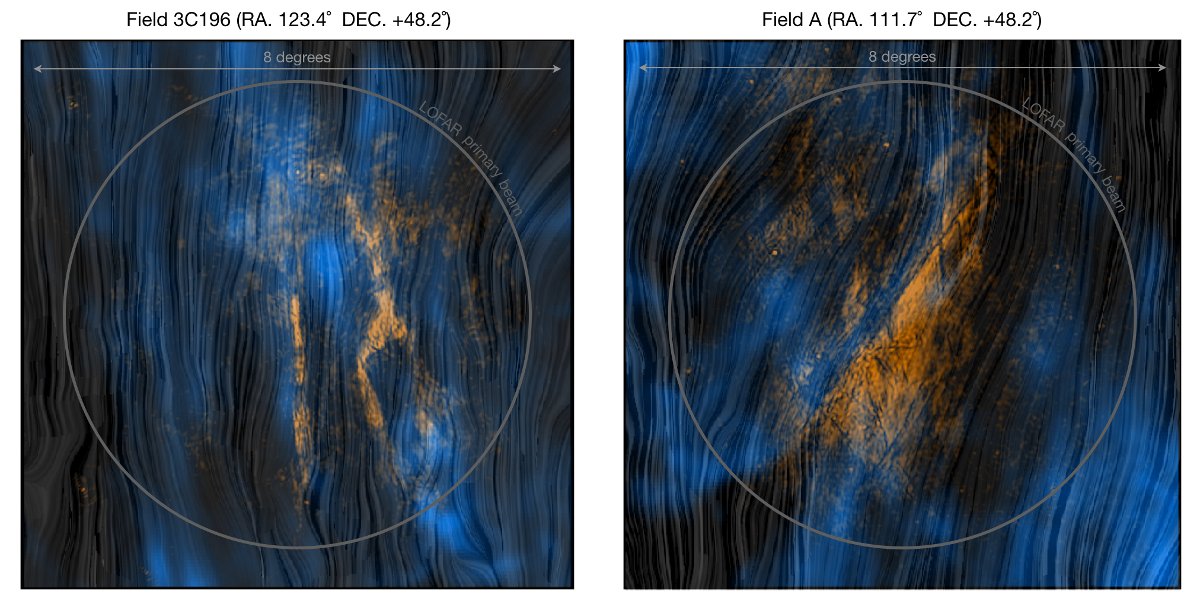}
\caption{Composite image showing the visual correlation of EBHIS HI 21cm brightness temperature ($T_{\rm b}(V_{\rm lsr})$, blue) with LOFAR HBA Faraday tomographic data ($\tilde{F}(\phi)$, orange) and \Planck\ magnetic-field orientations, $\chi_{\rm d}$, at 353 GHz (drapery pattern). All maps are shown in celestial coordinates. {\it Left panel}: $T_{\rm b}$ of Field 3C196  at $V_{\rm lsr} = -5$~km s$^{-1}$ and $\tilde{F}(\phi)$ at $\phi = 0\, {\rm rad\, m}^{-2}$. {\it Right panel}: $T_{\rm b}$ for Field A at $V_{\rm lsr} = +1.4$~km s$^{-1}$ and $\tilde{F}(\phi)$ at $\phi = -1.25\, {\rm rad\, m}^{-2}$. The gray circle in each panel traces the size of the primary LOFAR beam.}
\label{fig:LofarFields}
\end{figure*}

We analyzed the LOFAR observations of diffuse polarized emission detected in four fields at mid-to-high Galactic latitudes. These are Field 3C196 and three additional fields in its environs. Field 3C196 is centered on the very bright quasar 3C196, in one of the most diffuse interstellar windows of the northern sky \citet{Jelic2015}. We used Faraday depth data cubes (hereafter, $\tilde{F}({\rm RA,DEC,}\,\phi)$) -- or hyper-spectral cubes of polarized emission as a function of sky position and $\phi$ --  of four fields observed with the LOFAR high band antennas (HBAs, see Table~\ref{tab:fovs}).  
The $\tilde{F}({\rm RA,DEC,}\,\phi)$ cubes have an angular resolution of $\sim$ $4\,\arcmin$ and a Faraday depth resolution of $\sim1\,{\rm rad\, m^{-2}}$. The cubes cover a Faraday depth range between $-10$ and $+10\,{\rm rad\, m^{-2}}$, where most of the diffuse polarized emission is observed, in $\phi$ steps of $0.25\,{\rm  rad\, m^{-2}}$. The field of view of each cube is $\sim 64\,{\rm deg^2}$. The data for Field 3C196 are publicly available \citep{Jelic2015}\footnote{\url{http://cdsarc.u-strasbg.fr/viz-bin/qcat?J/A+A/583/A137}}. The data for the other three fields, Fields A, B, and C, are based on LOFAR data taken under the project code LC5\_11 (PI: V. Jeli\'c; {\color{blue} Turi\'c et al., in prep.}). The data were calibrated in a direction-independent manner and reduced in a similar way to what is described in \citep{Jelic2015}. The Faraday depth cubes were synthesized from the Stokes $Q$ and $U$ images of 170 frequency sub-bands, with a comparable noise level and frequency range between 115 MHz and 150 MHz and with a spectral resolution of 183 kHz.

\begin{table}[!h]
    \centering
    \begin{tabular}{|c|c|c|c|} \hline\hline
        & RA & DEC & \\  \hline 
         Field 3C196 & $123^{\circ}.4$ & $+48^{\circ}.2$ & \citet{Jelic2015} \\ \hline
         Field A & $111^{\circ}.7$ & $+48^{\circ}.2$ & \\ 
         Field B & $123^{\circ}.4$ & $+40^{\circ}.4$ & {\color{blue} Turi\'c et al., in prep.}\\ 
         Field C & $131^{\circ}.2$ & $+33^{\circ}.9$ &  \\ \hline\hline
    \end{tabular}
    \vspace{0.1cm} 
    \caption{Central celestial coordinates of the LOFAR fields of view studied in this work. All fields of view spanned an area of $\sim$64 deg$^2$.}
    \label{tab:fovs}
\end{table}

\subsection{EBHIS spectroscopic data at 21cm}\label{ssec:ebhis}
We used the publicly available spectroscopic data at 21 cm from the Effelsberg-Bonn HI Survey\footnote{\url{http://cdsarc.u-strasbg.fr/viz-bin/qcat?J/A+A/585/A41}} \citep[EBHIS,][]{Winkel2016} to trace the atomic medium toward our fields of view. The EBHIS survey covers the full northern sky (DEC $> -5^{\circ}$) at an angular resolution of $10.8\arcmin$, a spectral resolution of $1.44$ km s$^{-1}$ for a local-standard-of-rest velocity range of $|V_{\rm lsr}| \le 600$ km s$^{-1}$, and a brightness temperature noise level of $90$ mK. We used the EBHIS data at their nominal angular resolution. The data come in the form of position-position-velocity (PPV)
cubes, for which we considered both the brightness temperature ($T_{\rm b}$) as a function of $V_{\rm lsr}$ and the gas column density defined as: 
\begin{equation}\label{eq:M0}
     \frac{N_{\rm H}}{[\rm cm^{-2}]} = 1.82 \times 10^{18} \int \frac{T_{\rm b}(V_{\rm lsr})}{\rm [K]}\, \frac{{\rm d} V_{\rm lsr}}{\rm [km\, s^{-1}]}.
\end{equation}
This equation for $N_{\rm H}$ is valid under the assumption of optically thin HI emission \citep{Dickey1990}. In our case, the optically thin assumption is justified by the position of the LOFAR fields of view, which are among the most diffuse regions of the northern sky. 

\subsection{Planck polarization data}\label{ssec:planck}
We obtained the magnetic-field orientation from dust polarization using the publicly available \Planck\ PR3 polarization data\footnote{\url{http://www.cosmos.esa.int/web/planck/pla}} at 353 GHz \citep{PlanckIII2020}. At this frequency Faraday rotation is negligible. We used the Stokes $Q_{\rm d}$ and $U_{\rm d}$ maps in the \healpix\footnote{\url{http://healpix.sourceforge.net}} format of dust polarized emission (labeled with the subscript "d") smoothed at the angular resolution of $30\arcmin$ to increase the signal-to-noise ratio in polarization at the intermediate Galactic latitudes of the LOFAR fields of view. The smoothed maps were also rotated from Galactic to celestial coordinates before extracting cutouts of the fields in Table~\ref{tab:fovs}. From the cutout Stokes maps we obtained the polarization angle $\psi_{\rm d}= 0.5\,{\rm tan}^{-1}\left(-U_{\rm d},Q_{\rm d}\right)$, where the minus sign is needed to transform the {\healpix}-format maps into the International Astronomical Union (IAU) convention for $\psi_{\rm d}$, measured from the local direction to the north Galactic pole with increasing
positive values toward the east. We used the version of the inverse tangent function with two signed arguments to solve the $\pi$ ambiguity in the definition of $\psi_{\rm d}$, as it corresponds to orientations and not to directions. The magnetic-field orientation, $\chi_{\rm d}$, was finally obtained as $\chi_{\rm d} = \psi_{\rm d} + \pi/2$. 

\section{Statistical correlation between LOFAR and EBHIS}\label{sec:methods}

Here we describe the statistical correlation between the LOFAR and EBHIS datasets. 

\subsection{A visual inspection of the data}\label{ssec:visual}

As an example, in Fig.~\ref{fig:LofarFields}, we show the $T_{\rm b}$ (in blue) of two fields of view (Field 3C196 and Field A) as well as their polarized emission from LOFAR tomographic data (in orange). This figure highlights the visual correspondence of morphological structures in both datasets that also correlate with the plane-of-the-sky orientation of the magnetic field encoded in the \Planck\ data. Figure~\ref{fig:LofarFields} shows $\chi_{\rm d}$ as drapery pattern obtained using the line integral convolution \citep[LIC,][]{Cabral1993} technique. Within the primary beam of LOFAR (gray circle in both panels), the association between $T_{\rm b} (-5\, {\rm km\, s^{-1}})$ and $\tilde{F}(0\, {\rm rad\, m}^{-2})$ in the case of Field 3C196, and $T_{\rm b} (+1.4\, {\rm km\, s^{-1}})$ and $\tilde{F}(-1.25\, {\rm rad \,m}^{-2})$ in the case of Field A, is striking. Not only does the structure of $T_{\rm b}$ follow that of $\tilde{F}({\rm RA,DEC,}\,\phi)$, but both datasets also show a correlation with the mean orientation of $\chi_{\rm d}$ and with localized bends of the magnetic-field lines\footnote{We note that in Fig.~\ref{fig:LofarFields} we only show the emission for one velocity channel in the EBHIS data and for one Faraday depth in the LOFAR data. In both cases, the emission at a specific channel may not cover the entire field of view in the same way. Moreover, in the LOFAR case, the primary beam represented by the gray circle strongly attenuates the emission at the edges compared to the EBHIS data.}.      
\begin{figure*}[!ht]
\centering
\includegraphics[width=.9\textwidth]{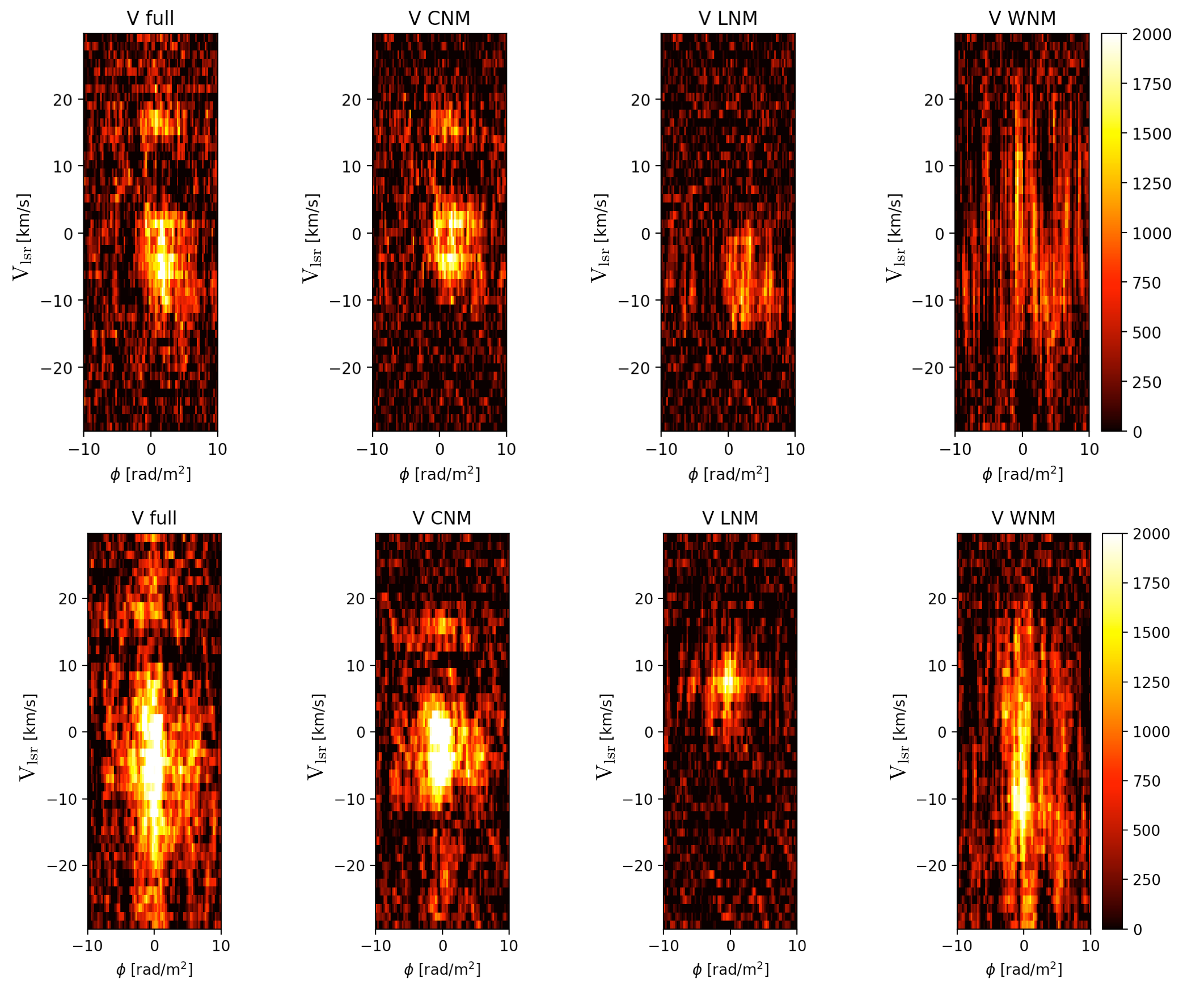}
\caption{Projected Rayleigh statistics $V$ using HOG between LOFAR and EBHIS data for Field 3C196 (top) and Field A (bottom). From left to right, we consider the full EBHIS dataset and the individual CNM, LNM, and WNM components obtained with {\tt ROHSA}. The dynamic range of V shown by the color bar is the same for all eight panels.}
\label{fig:hog}
\end{figure*}

We recall that the correlation between the magnetic-field structure and the HI emission seen in a thin ($\sim$1 km s$^{-1}$) velocity channel is usually
associated with the CNM  \citep{Clark2019,Peek2019}. Nevertheless, to better understand the physical conditions that give rise to the observed correlation (HI, dust polarization, and Faraday tomography), it is necessary to study in greater detail the multiphase structure of the HI gas in each field.

\subsection{Measuring the correlation in the multiphase HI}\label{ssec:HIgas}

We investigated the radiative condensation of the diffuse WNM in thermally unstable LNM and stable CNM based on HI emission from EBHIS PPV cubes. In order to disentangle the multiphase-HI-blended signal along the line of sight, we relied on the Gaussian approximation of spectroscopic lines \citep[e.g.,][]{Matthews1957,Dieter1965,Takakubo1966,Mebold1972,Haud2007,Kalberla2018}, differing from warm to cold HI phases according to their observed standard deviations, $\sigma_{\rm Gauss}$. We made use of one recent approach, described in \citet{Marchal2019}, called {\it Regularized Optimization for Hyper-Spectral Analysis} ({\tt ROHSA})\footnote{\url{https://github.com/antoinemarchal/ROHSA}}. {\tt ROHSA} performs a regression analysis using a regularized nonlinear least-square criterion. The novelty of {\tt ROHSA} is the search for spatial coherence in the determination of the parameters of the Gaussian spectral decomposition (see Appendix~\ref{app:rohsa} for more details on {\tt ROHSA} and on its validation with respect to this work).

As shown in Fig.~\ref{fig:rohsa}, we were able to build consistent three-HI-phase models for our fields of view with an accuracy of a few percent. We used thresholds on $\sigma_{\rm Gauss}$ of 3 km s$^{-1}$ and 6 km s$^{-1}$ to discriminate between CNM and LNM and between LNM and WNM, respectively. Results did not change for 1 km s$^{-1}$ variations of these thresholds. {\tt ROHSA} provided us with four PPV cubes per field of view (full, CNM, LNM, and WNM) that we could statistically compare with the corresponding cubes of $\tilde{F}({\rm RA,DEC,}\,\phi)$. 

The statistical analysis of the correlation between  HI emission and Faraday tomographic cubes was made using the histogram of oriented gradients (HOG)\footnote{\url{http://github.com/solerjuan/astrohog}}, a tool presented in \citet{Soler2019} as a new metric for systematic characterizations of PPV hyper-spectral cubes of atomic and molecular gas. The basic principle of HOG is that it estimates the spatial correlation (morphological alignment) between two images -- or maps -- assuming that the local appearance and shape of an image can be fully characterized by the distribution of its local intensity gradients or edge directions (see Appendix~\ref{app:hog}).

We applied HOG to the EBHIS and LOFAR data at the EBHIS angular and pixel resolution. We computed the relative orientation between the directions of the local gradients of $T_{\rm b} (V_{\rm lsr})$ and $\tilde{F}(\phi)$ at each velocity channel and Faraday depth. As suggested by \citet{Soler2019}, to evaluate the correlation we used the HOG output parameter defined as the projected Rayleigh statistics ($V$, see Eq.~\ref{eq:Vhog}), which represents the likelihood that the gradients of two emission maps ($T_{\rm b} (V_{\rm lsr})$ and $\tilde{F}(\phi)$ in our case) are mostly parallel. Like \citet{Soler2019}, we also noticed that it is not possible to draw conclusions from the values of $V$ alone, but its statistical significance can be assessed by comparing a given $V$ value to others obtained in maps with similar statistical properties. 
Using HOG, we were able to build $V$ maps as a function of the ranges of $V_{\rm lsr}$ and $\phi$ that were detected by EBHIS and LOFAR, respectively. 

In Fig.~\ref{fig:hog}, we show $V$ maps of Field 3C196 (top row) and of Field A (bottom row) as a function of $V_{\rm lsr}$ between $-30$ and $+30$ km s$^{-1}$ and of $\phi$ between $-10$ and $+10$ rad m$^{-2}$, that is to say, the intervals of velocity and Faraday depth at which most of the emission is observed. The $V$ maps are presented for the full HI EBHIS PPV cube and separately for the CNM, LNM, and WNM PPV cubes. In both fields of view, a clear morphological correlation between LOFAR and EBHIS data is found based on the HOG results.

Two distinct bright spots in both rows can be identified in the full PPV cube diagram on the left at $V_{\rm lsr}\approx 15$ and $-5$ km s$^{-1}$. These two bright $V$-spots cover a broad range of $|\phi|$, from a few up to $10$ rad m$^{-2}$; the range is mostly at positive $\phi$ for Field 3C196, while it is more symmetric around $0$ rad m$^{-2}$ in the case of Field A. 

As shown by the $V$ maps derived from each HI phase, most of the total correlation can be retrieved in the CNM component, although the contributions from the LNM or WNM are not negligible. The $V$-bright spot at negative velocities is apparent in the LNM component of Field 3C196 as well. The WNM shows a dimmer but more extended correlation, probably due to the large widths of the spectral lines, which does not resemble noise. The effect of noise in the correlation can be seen in the granularity of the top part of the LNM diagram. In the case of Field A, the LNM component shows a bright spot in $V$ at $\sim$7 km s$^{-1}$, while the WNM component has brighter and more diffuse $V$ values at negative velocities, which are also observed in the case of the full PPV cube on the left.

The correlation based on the $V$ parameter is shared among all three HI phases. The CNM component is the dominant one in determining the link between the LOFAR and EBHIS data. In some cases, however, the association of the LOFAR data with all three phases at specific velocities (see, for instance, the bright $V$ spot at negative velocities for Field 3C196 in Fig.~\ref{fig:hog}) may suggest a real physical mixture of the different HI phases, possibly a sign of phase transition at play. 

\begin{figure}[!t]
\centering
\includegraphics[width=0.49\textwidth]{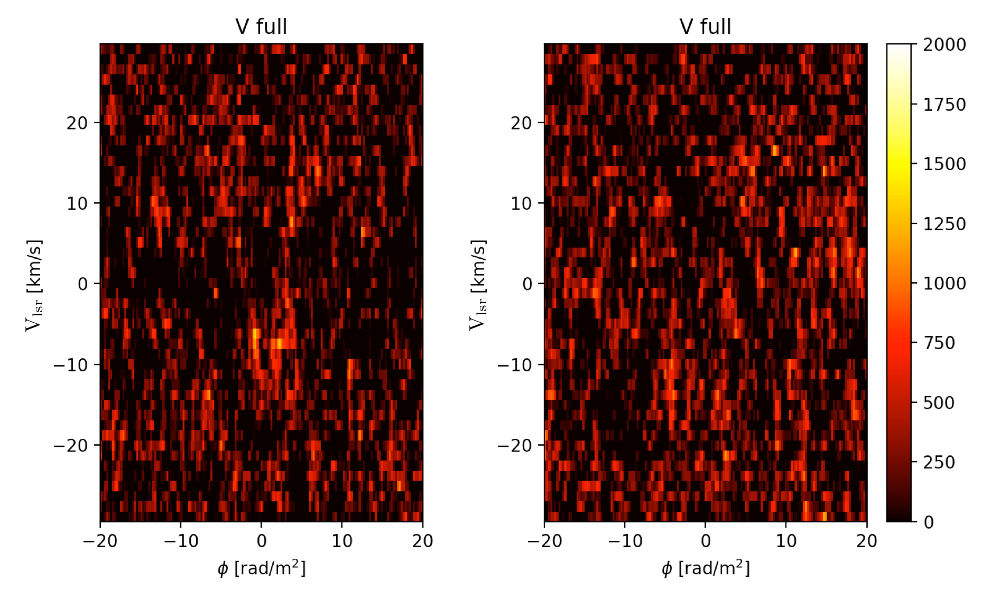}
\caption{Same as in Fig.~\ref{fig:hog} for two additional fields of view, Field B (left) and Field C (right).}
\label{fig:hogBC}
\end{figure}

Regarding Fields B and C, we find no morphological correlation (or almost none) between the LOFAR and EBHIS data using HOG as, illustrated in Fig.~\ref{fig:hogBC}. We also explored a larger range of $\phi$, between $-20$ and $+20$ rad m$^{-2}$. Field B suggests a weak enhancement of $V$ at the same location shown by Field 3C196 and Field A (at $V_{\rm lsr}\approx -5$ km s$^{-1}$). In the case of Field C, all values of $V$ are consistent with noise. As  will be discussed in more detail by {\color{blue} Turi\'c et al. (in prep.)}, the polarized emission in Field C is significantly different from the other fields of view. It is generally weaker and with a patchier morphology. We cannot conclude that Fields B and C intrinsically lack correlation. As we discuss in Appendix~\ref{app:SNR}, a possible explanation for the observed difference among the four fields of view is likely the combination of the low signal-to-noise ratio in LOFAR polarization with the low efficiency of HOG in detecting alignment in sparse, patchy images.

\section{On the correlation between LOFAR and HI data}\label{sec:hog}

\begin{figure}[!t]
\centering
\includegraphics[width=0.50\textwidth]{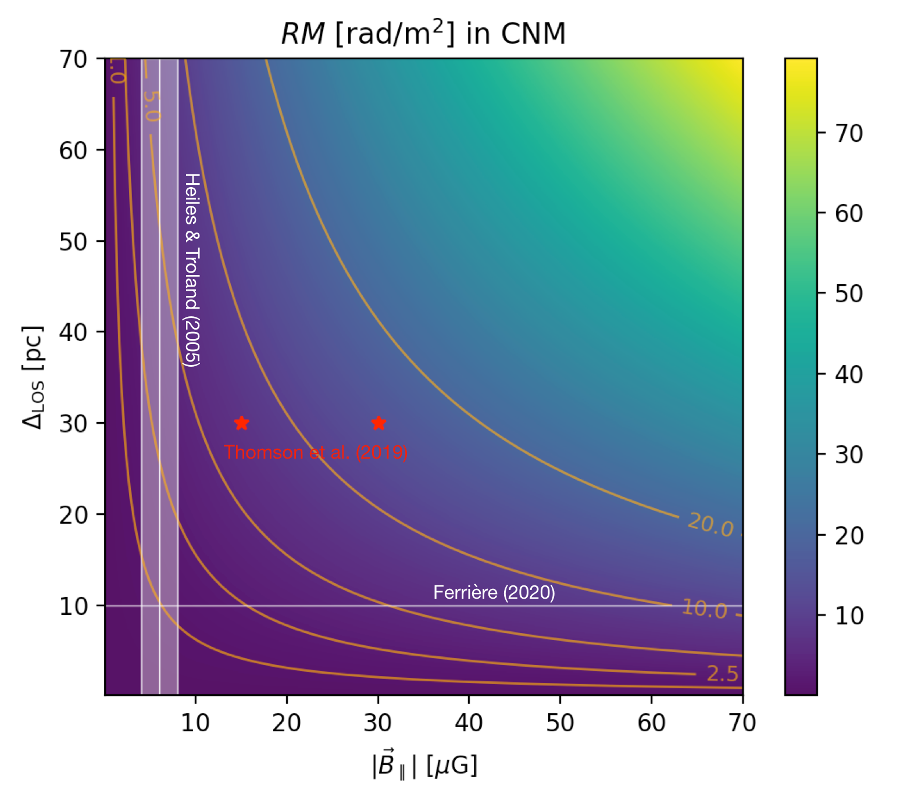}
\caption{Expected values of $RM$ for a CNM cloud given a typical electron density for the CNM of $n_e = 0.02$ cm$^{-3}$ \citep{Ferriere2020}. $RM$ is shown as a function of the strength of magnetic-field component parallel to the line of sight, $B_{\parallel}$, and of the line-of-sight thickness of the cloud, $\Delta_{\rm LOS}$. Orange contours show $RM$ values of 1, 2.5, 5, 10, and 20 rad m$^{-2}$. Vertical white lines and shades correspond to the median and standard deviation of $B_{\parallel}$ found in the CNM \citep{Heiles2005}. The horizontal white line indicates the standard size of a CNM cloud \citep{Ferriere2020}. Two red stars mark the values of $RM$ found in \citet{Thomson2019} and associated with CNM clouds.}
\label{fig:RMcnm}
\end{figure}
Focusing on Fields 3C196 and A, the correlation between all HI phases, notably the CNM, with the Faraday rotation detected by LOFAR in the magneto-ionic ISM is a puzzling result. Faraday rotation is expected to be effective in highly ionized interstellar gas (WIM) rather than in the atomic medium. 

The observed correlation may be related either to the structure of the intervening ISM that produces Faraday rotation or to the synchrotron-emitting regions along the sight line -- or to a mixture of the two processes.
Regarding the first case, we show in Fig.~\ref{fig:RMcnm} the expected values of rotation measure ($RM$, see Eq.~\ref{eq:RM}) analytically derived for a single CNM cloud along the line of sight as a function of its line-of-sight thickness ($\Delta_{\rm LOS}$) and magnetic-field strength ($B_\parallel$), given a standard thermal electron density for the CNM of $\sim$0.02 cm$^{-3}$ \citep{Ferriere2020}. As indicated by the white shades and lines in the figure, standard physical sizes of CNM clouds ($\sim$10 pc) and magnetic-field strengths ($6 \pm 2\, \mu$G) measured with Zeeman splitting at 21cm \citep{Heiles2005} are not capable of producing $RM$ values larger than a few rad m$^{-2}$. Therefore, the values observed in this study (up to 10 rad m$^{-2}$) appear significantly unlikely for the diffuse HI gas unless one considers either a large line-of-sight thickness ($\sim$60 pc) or a large magnetic-field strength up to $\sim$25 $\mu$G. \citet{Thomson2019} also interpreted their tomographic data with the Parkes telescope in terms of Faraday-rotating CNM structures along the line of sight. They claimed to have measured the Faraday signature of two CNM clouds at roughly $\phi\approx -7$ and $14$ rad m$^{-2}$, which correspond to magnetic-field strengths of $15$ and $30$ $\mu$G (see red stars in Fig.~\ref{fig:RMcnm}). Such magnetic-field strengths are also unusually strong for the CNM. These high values may be explained by extreme nonstandard physical conditions that may trigger phase transition in the multiphase ISM.
Large-scale compressions caused by interstellar shocks, which are identified in theory as key mechanisms for CNM formation \citep[e.g.,][]{Inutsuka2015,Inoue2016}, could possibly account for the large magnetic-field strengths caused by supersonic gas motions in frozen-in interstellar magnetic fields. 

This interpretation would be supported by the location of our four LOFAR fields of view in the environs (in projection) of known loop-like structures that have been interpreted as being the result of interstellar shocks caused by either supernova remnants (SNRs) or rain-falling high-velocity atomic clouds in the Galactic plane. Some of these structures can be identified as the radio Loop III \citep{Quigley1965}, the North Celestial Loop \citep{Meyerdierks1991}, and the Ursa Major Circle \citep{Bracco2020}. Establishing whether the LOFAR fields are associated with these specific loops or to other shell-like structures seen in HI emission toward the same region \citep[e.g.,][]{Heiles1979,Ehlerova2013} is beyond the scope of this work. An upcoming paper will possibly clarify this interpretation as it will focus on the analysis of the LOFAR Two-meter Sky Survey \citep[LoTSS,][]{Shimwell2017,Shimwell2019} tomographic data of the entire Loop III region ({\color{blue} Erceg et al., in prep.}).

Interstellar shocks at the origin of the correlation between the Faraday rotation observed with LOFAR and the neutral structures along the sight line observed with interstellar dust extinction were also proposed by \citet{vanEck2017}. In the absence of known SNRs that may have generated the shock waves, the authors suggested other valuable scenarios that could also apply to our case. In particular, \citet{vanEck2017} noted that the correlation might be caused by complete depolarization in the WIM at LOFAR frequencies, leaving only a visible polarized signal coming from the neutral clouds. If this were true, the observed correlation would be associated with synchrotron emission from neutral clouds rather than with Faraday rotation. Even though the ionized medium would not produce Faraday depth peaks, it would Faraday-rotate any background signal, possibly shifting the intrinsic $\phi$ of a neutral cloud by a few rad m$^{-2}$ to the $\phi$ values observed in this letter. More work is needed to resolve the ambiguity between Faraday rotation and synchrotron emission from neutral clouds in order to provide the most likely interpretation of our results, and in particular the dominant role of the CNM over the other HI phases. 

\section{Conclusions}\label{sec:summary}

In this letter, we presented the first statistical analysis of the correlation between the magneto-ionic medium, probed by the LOFAR telescope through Faraday tomography, and the diffuse and multiphase neutral ISM traced by the HI spectroscopic data from EBHIS. We analyzed four distinct fields of view and found that the correlation between the two datasets can be striking. Two fields of view, out of the four, show an unquestionable correlation between the LOFAR and EBHIS data, in particular between Faraday tomographic data and the CNM gas. The correlation with the LNM and WNM phases is not negligible. We discussed possible mechanisms that may have generated the observed correlation, from the Faraday rotation of compressed neutral regions to the synchrotron emission of CNM clouds. However, future work is needed to pinpoint the most likely scenario. In any case, our results showed how the complex structure of the ionized medium seen by LOFAR tomographic data is tightly related to the multiphase and magnetized neutral ISM.

\begin{acknowledgements}
We thank the valuable comments of the anonymous referee that improved the presentation of our work. We are thankful to Tim Robishaw for a detailed revision of the manuscript. AB acknowledges the support from the European Union’s Horizon 2020 research and innovation program under the Marie Skłodowska-Curie Grant agreement No. 843008 (MUSICA). AB also thanks JUR for helping keep mental sanity in these crazy times (\url{http://youtu.be/2h5hQUqYz0w}). VJ, LT and AE acknowledge support by the Croatian Science Foundation for the project IP-2018-01-2889 (LowFreqCRO). VJ and LT also acknowledge support by the Croatian Science Foundation for the project DOK-2018-09-9169. This paper is based in part on data obtained with the International LOFAR Telescope (ILT) under project code LC\_008. LOFAR (van Haarlem et al. 2013) is the Low Frequency Array designed and constructed by ASTRON. It has observing, data processing, and data storage facilities in several countries, that are owned by various parties (each with their own funding sources), and that are collectively operated by the ILT foundation under a joint scientific policy. The ILT resources have benefitted from the following recent major funding sources: CNRS-INSU, Observatoire de Paris and Université d'Orléans, France; BMBF, MIWF-NRW, MPG, Germany; Science Foundation Ireland (SFI), Department of Business, Enterprise and Innovation (DBEI), Ireland; NWO, The Netherlands; The Science and Technology Facilities Council, UK; Ministry of Science and Higher Education, Poland.
\end{acknowledgements}

\bibliographystyle{aa}
\bibliography{aanda.bbl}

\appendix

\section{Formalism of Faraday rotation}\label{app:farad}

Faraday rotation is a wavelength-dependent ($\lambda$) process that empirically affects the measure of the plane-of-the-sky orientation of the magnetic field, $\chi$, inferred from synchrotron polarization as:
\begin{equation}\label{eq:rmemp}
RM = \frac{{\rm d} \chi}{{\rm d}\lambda^2}, 
\end{equation}
where $RM$ is defined as the rotation measure in units of rad m$^{-2}$. This amount of rotation is physically expected to depend on two main parameters, namely the free-electron density, $n_e$, and the magnetic-field strength along the line of sight as 
\begin{equation}\label{eq:RM}
    \frac{RM}{[\rm rad/m^2]} = 0.81\, \int^{0}_{\rm source} n_e\, \vec{B} \cdot {\rm d} \vec{l},
\end{equation}
where $n_e$ is in units of cm$^{-3}$, $|\vec{B}_{\parallel}|$ is in units of $\mu$G, and the distance to the synchrotron emitting source is measured in parsecs. The value of $RM$ can be positive or negative depending on the direction of $\vec{B}$, toward the observer or away from the observer, respectively. In Sect.~\ref{sec:hog}, we analytically estimate $RM$ for a CNM cloud as $\sim 0.81\, n_e B_{\parallel}\Delta_{\rm LOS}$, where $\Delta_{\rm LOS}$ represents the line-of-sight thickness of the CNM cloud.  

Faraday tomography introduces a novel independent variable, namely Faraday depth, $\phi$, which has the same units as $RM$. However, it represents the specific amount of Faraday rotation that the synchrotron polarization experiences within any distinct parcel of ionized gas along the line of sight. 
Similarly to $RM$, $\phi$ is defined as 
\begin{equation}\label{eq:FD}
    \frac{\phi(\vec{l})}{[\rm rad/m^2]} = 0.81\, \int^{L_0}_{L_0 + \Delta L} n_e\, \vec{B} \cdot {\rm d} \vec{l}, 
\end{equation}
where the line-of-sight integration now involves a confined region of magneto-ionic ISM, namely $\Delta L$ at the distance $L_{0}$.  In other words, $\phi$ is equivalent to $RM$ in the limit of one single magnetized cloud of free electrons Faraday-rotating a background synchrotron emission. As radio-polarimetric observations of linear polarization consist of measuring the Stokes parameters $Q$ and $U$ as a function of $\lambda$, it is possible to define the complex polarization vector, $P(\lambda^2) = Q(\lambda^2) + i U(\lambda^2)$, and decompose it into its distinct components in Faraday space. Similarly to the decomposition of a signal in Fourier space, one can introduce
\begin{equation}\label{eq:FT}
    F(\phi) = \frac{1}{\pi} \int_{-\infty}^{+\infty} P(\lambda^2)e^{-2i\phi\lambda^2} \,{\rm d}\lambda^2, 
\end{equation}
where $F(\phi)$ now represents the Fourier conjugate function to $P(\lambda^2)$, and its modulus is $\tilde{F}(\phi) = \sqrt{Q(\phi)^2+U(\phi)^2}$ in units of polarized intensity. More details about the discretization of the $\lambda^2$-space in the actual computation of $F(\phi)$ are available in \citet{Brentjens2005}. 

\section{HI decomposition with {\tt ROHSA}}\label{app:rohsa}
\begin{figure}[!ht]
\centering
\includegraphics[width=0.9\linewidth]{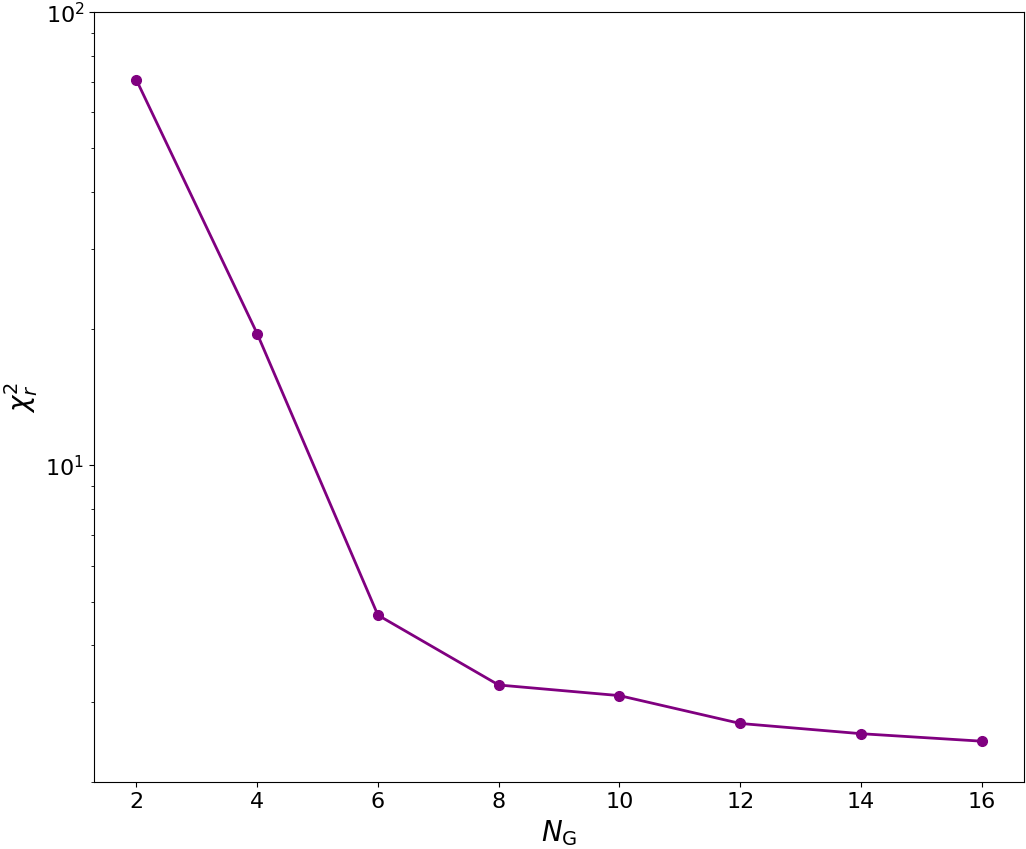}
\caption{Reduced chi-square $\chi^2_{\rm r}$ of {\tt ROHSA} solutions as a function of the number of Gaussian $N_{\rm G}$ used to decompose EBHIS data centered on Field 3C196.
}
\label{fig:chi2}
\end{figure}

\begin{figure}[!ht]
\centering
\includegraphics[width=0.9\linewidth]{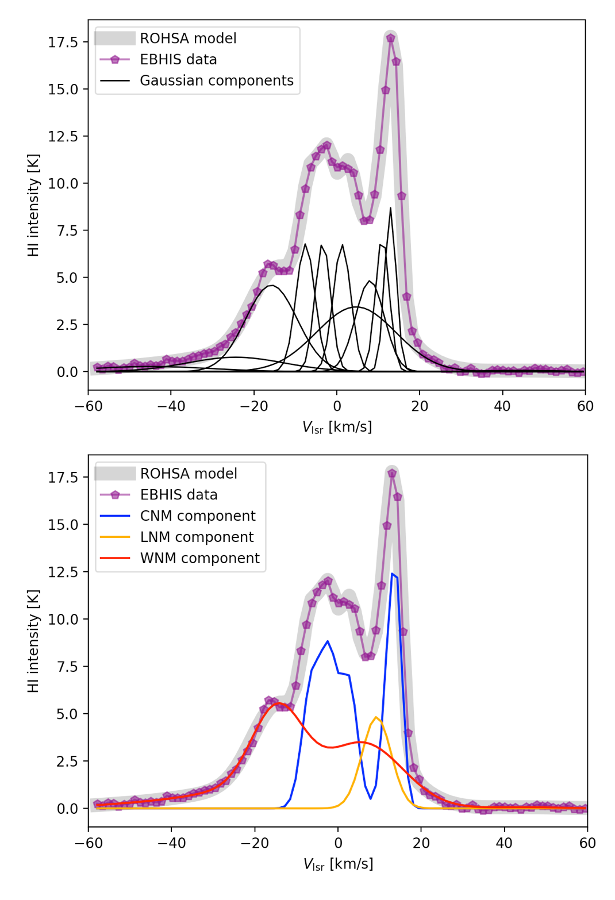}
\caption{Example of a {\tt ROHSA} decomposition of the HI spectrum of EBHIS toward the central coordinates of Field A. {\it Top panel}: EBHIS HI spectrum shown in purple, Gaussian components in black, and the reconstructed modeled spectrum from {\tt ROHSA} in gray. {\it Bottom panel}: Same line of sight as shown in the top panel, with the addition of cumulative contributions of CNM (blue), LNM (yellow), and WNM (red).}
\label{fig:rohsasp}
\end{figure}
\begin{figure}[!ht]
\centering
\includegraphics[width=0.99\linewidth]{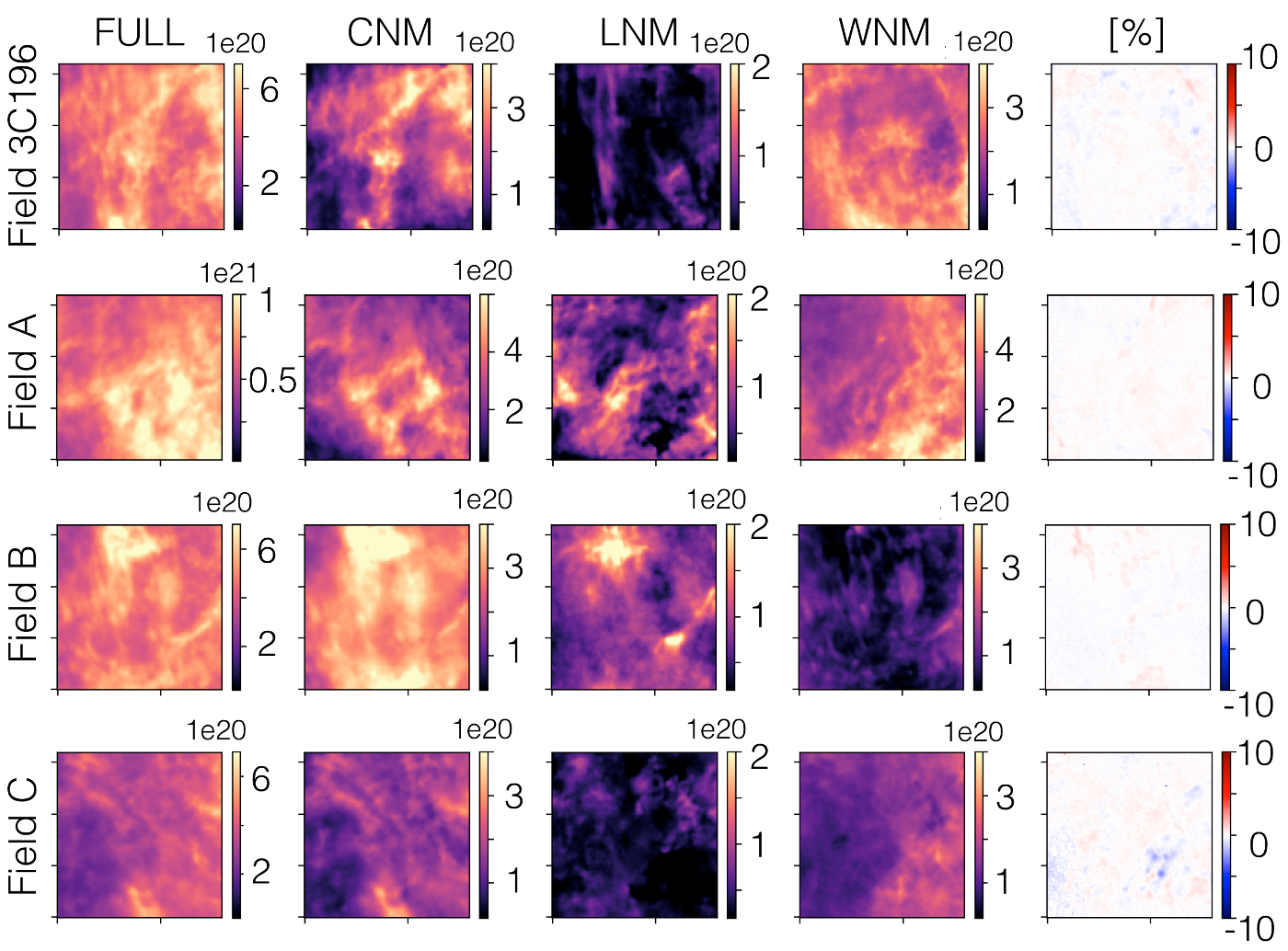}
\caption{Maps of the HI column density, $N_{\rm H}$, in units of cm$^{-2}$, of the PPV cubes from all fields of view. From top to bottom, we show Fields 3C196, A, B, and C. The $N_{\rm H}$ is computed between $-20 < V_{\rm lsr}/ [\rm km\, s^{-1}] < +20$. From left to right, $N_{\rm H}$ is shown for the full EBHIS dataset and for the CNM, LNM, and WNM components. The last column on the right shows the residual map between $N_{\rm H}$(full) and $N_{\rm H}$(CNM+LNM+WNM).   
}
\label{fig:rohsa}
\end{figure}


{\tt ROHSA} performs a regression analysis on the whole PPV cube simultaneously to decompose the signal into a sum of $N_{\rm G}$ spatially coherent Gaussians parametrized by three 2D fields: amplitude, mean velocity and velocity dispersion ($\bm{a}_n$,\,$\bm{\mu}_n$,\,$\bm{\sigma}_n$).     
This is achieved by adding regularization terms (i.e., a Laplacian filtering) in the cost function that penalizes the small spatial frequencies of each parameter maps. 
In addition, {\tt ROHSA} minimizes the variance of $\bm{\sigma}_n$ to enable the multiphase separation.
The magnitude of these four energy terms is controlled by the hyper-parameters $\lambda_{\bm{a}}$, $\lambda_{\bm{\mu}}$, $\lambda_{\bm{\sigma}}$, and $\lambda_{\bm{\sigma}}'$.
Finally, the initialization of the Gaussian parameters is done in {\tt ROHSA} via a multi-resolution process from a coarse to a fine grid.
Further details are available in Sect.~2.4 of \cite{Marchal2019}.

We decomposed EBHIS PPV cubes centered on the four LOFAR fields analyzed here using {\tt ROHSA} user-parameters $N_{\rm G}=12$ and $\lambda_{\bm{a}} = \lambda_{\bm{\mu}} = \lambda_{\bm{\sigma}} = \lambda_{\bm{\sigma}}' = 400$. 
The four hyper-parameters were empirically chosen to ensure that each Gaussian has spatially coherent parameter maps and can be classified as CNM ($\left<\bm{\sigma}_n\right> < 3$ km\,s$^{-1}$), LNM (3 km\,s$^{-1} \le \left< \bm{\sigma}_n\right> < 6$ km\,s$^{-1}$) or WNM ($\left<\bm{\sigma}_n\right> \ge 6$ km\,s$^{-1}$) based on their mean velocity dispersion.
The number of Gaussians was chosen to converge toward a noise-dominated residual. To achieve this, and to assess the quality of the decomposition, we fixed the four hyper-parameters and varied $N_{\rm G}$ from two to 16. Figure~\ref{fig:chi2} shows the reduced chi-square $\chi^2_{\rm r}$ as a function of $N_{\rm G}$ for Field 3C196. The $\chi^2_{\rm r}$ converges toward $\sim2.5$ for $N_{\rm G} > 10$, showing that at least 12 Gaussians are needed to encode the signal.
Building on this, we kept the decomposition parametrized with $N_{\rm G}=12$. 
The impact of varying $N_{\rm G}$ from 12 to 16 on the final results presented in the letter (i.e., the LOFAR EBHIS spatial and spectral correlation) will be discussed in Appendix~\ref{app:impact}.

Figure~\ref{fig:rohsasp} shows an example of the three-phase model obtained with {\tt ROHSA}, parametrized with $N_{\rm G}=12$, toward the central coordinates of Field A. Data are shown in purple in the top panel, individual Gaussians are shown in black, and the reconstructed modeled HI spectrum is shown in gray. In the bottom panel of the same figure, we show the modeled spectrum as decomposed in each HI phase.

The consistent solution for all fields is shown in Fig.~\ref{fig:rohsa}, where we show maps of $N_{\rm H}$ (see Eq.~\ref{eq:M0} integrated in the range $-20 < V_{\rm lsr}/ [\rm km\, s^{-1}] < +20$) corresponding to each HI phase. Residual maps between {\tt ROHSA} outputs and full PPV cubes are also shown with an accuracy on the order of a few percent in the right-hand column. We notice that the column density in Field A is larger as it is the closest field of view to the Galactic plane.  

\section{The use of HOG}\label{app:hog}

In our analysis we made use of HOG to quantify the morphological alignment between the EBHIS HI brightness temperature ($T_{\rm b}$) at each velocity channel $l$ and the LOFAR polarized emission ($\tilde{F}$) per Faraday depth channel $m$ in the tomographic cubes. We used the projected Rayleigh statistics ($V$ in \citet{Soler2019}) as a measure of the alignment between the local gradients of $T_{\rm b}$ and $\tilde{F}$, defined as: 
\begin{equation}\label{eq:Vhog}
V_{lm} = \frac{\sum_{ij}w_{ij,lm}\cos{2\alpha_{ij,lm}}}{\sqrt{\sum_{ij}(w_{ij,lm})^2/2}},
\end{equation}
where $w_{ij,lm}$ is the statistical weight assigned to the relative orientation angle between the two gradients, $\alpha_{ij,lm}$ at the $ij$ position in the maps. As we were only interested in the alignment, we considered positive values of $V$.

\subsection{The impact of varying $N_{\rm G}$ on the LOFAR-EBHIS correlation}\label{app:impact}
Here we evaluate the robustness of our results with respect to the choice of {\tt ROHSA} user-parameters and the use of HOG. We checked that the choice of $N_{\rm G}$ for $\chi^2_{\rm r} \sim 2.5$ (see Appendix~\ref{app:rohsa}) was not a critical parameter. 
\begin{figure}[!ht]
\centering
\includegraphics[width=0.99\linewidth]{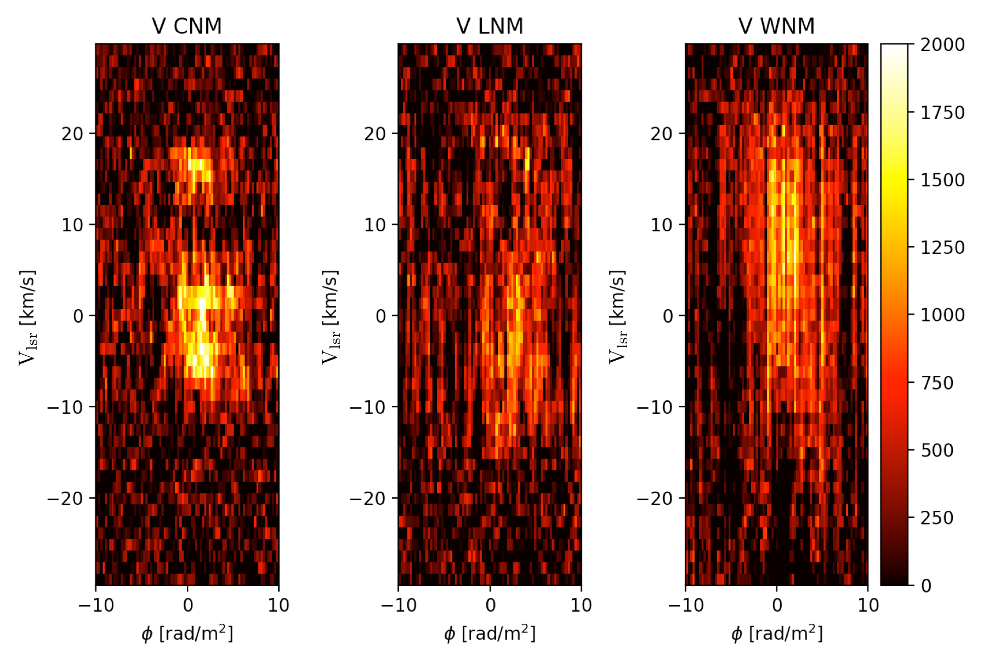}
\caption{Same as in the top row of Fig.~\ref{fig:hog} but with $N_{\rm G}=16$ instead of $N_{\rm G}=12$.}
\label{fig:16gauss}
\end{figure}

As Fig.~\ref{fig:16gauss} shows, we repeated the HOG analysis for Field 3C196 (see also Sect.~\ref{ssec:HIgas}) using $N_{\rm G}=16$ instead of $N_{\rm G}=12$ without noticing any significant change in our results (see the comparison with the top row in Fig.~\ref{fig:hog}). Although the $V$-patterns of the warmer phases are slightly different, the strong correlation between LOFAR and the CNM from EBHIS is preserved, as is the overall correlation with the LNM and the WNM.

\subsection{HOG efficiency with low signal to noise in polarization}\label{app:SNR}

\begin{figure}[!ht]
\centering
\includegraphics[width=0.49\textwidth]{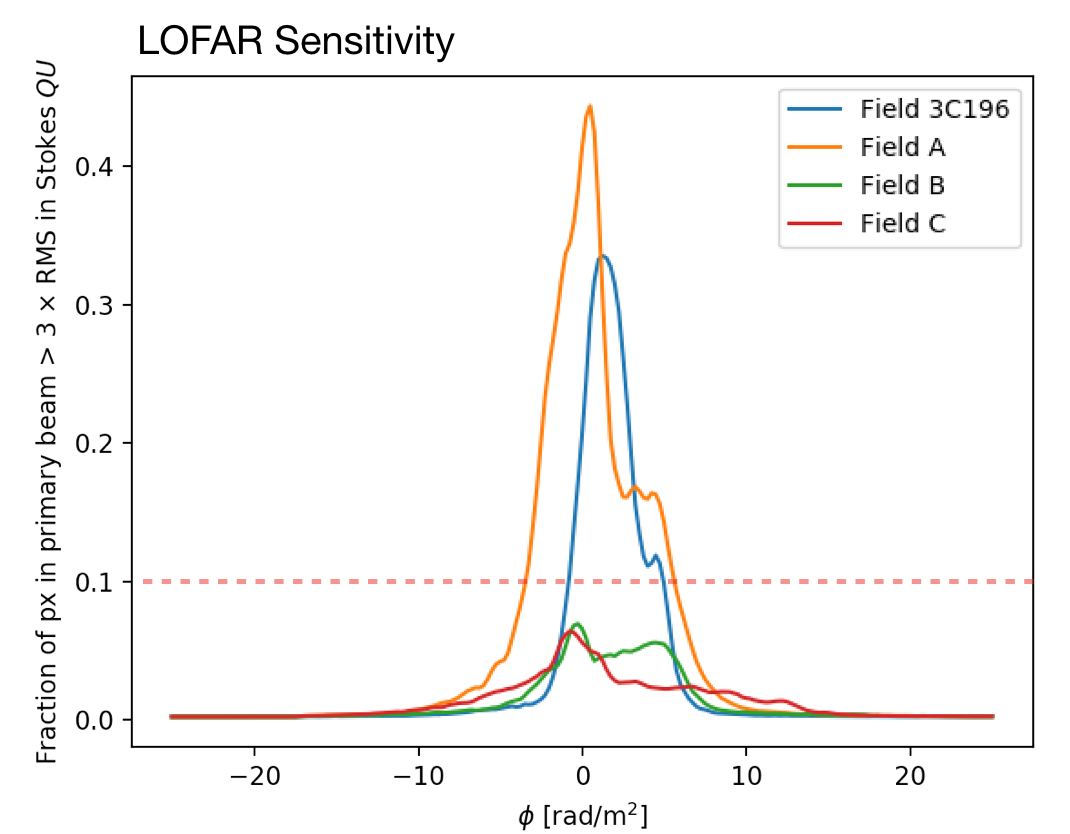}
\caption{Variation of LOFAR sensitivity per Faraday depth channel. The fraction of pixels with polarized intensity greater than three times the RMS value in both Stokes $Q$ and $U$ is shown. The dashed horizontal line represents a fraction of 10\%.}
\label{fig:lofsens}
\end{figure}

Fields B and C do not show a correlation between LOFAR and EBHIS data as shown in Fig.~\ref{fig:hogBC}. We believe that this is mainly caused by a low signal-to-noise ratio in polarization in the two fields. In Fig.~\ref{fig:lofsens}, we show the fraction of pixels in each $\tilde{F}$ map with at least three times the root-mean-square (RMS) value both in Stokes $Q$ and $U$ (on the order of $\sim$130 $\mu$Jy/PSF/RMSF) in the LOFAR data as a function of $\phi$. Fields B and C are both characterized by a low fraction of pixels ($< 10\,\%$) with high signal to noise across the full range of $\phi$.

\begin{figure}[!ht]
\centering
\includegraphics[width=0.49\textwidth]{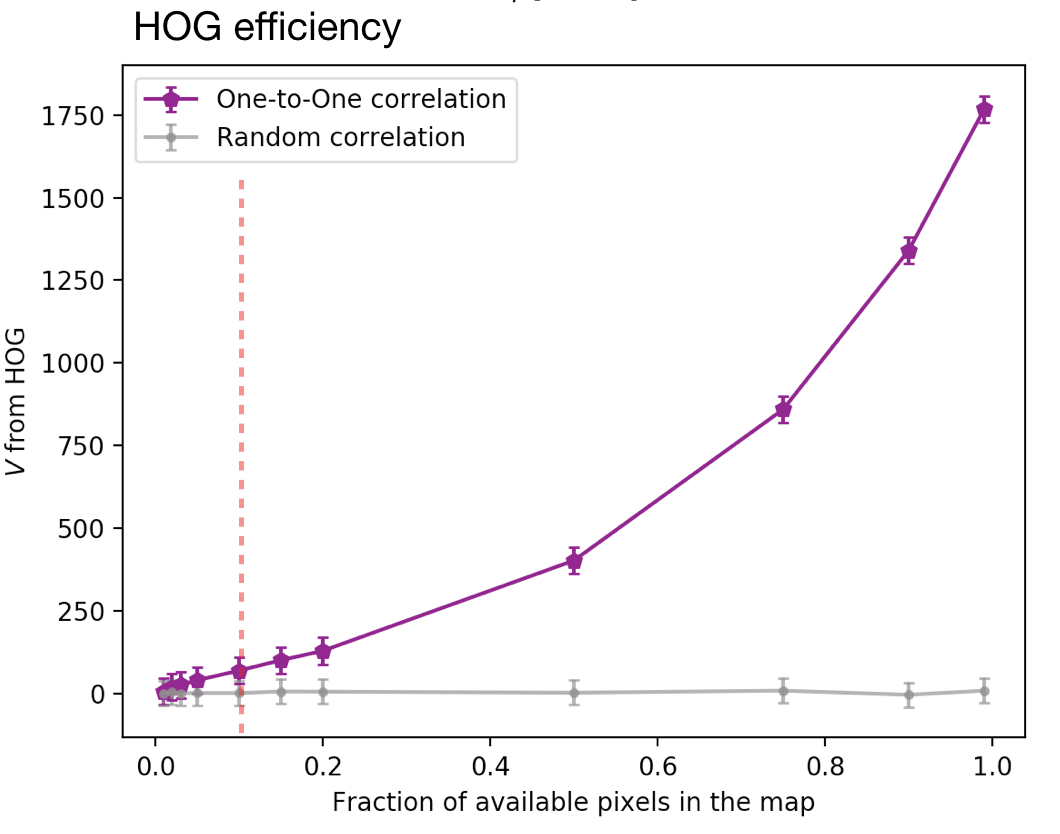}
\caption{HOG efficiency in finding morphological correlations in sparsely sampled images. The dashed vertical line represents 10\% of the available pixels in a given map.  
}
\label{fig:hogeff}
\end{figure}

These pixels are unevenly distributed on the maps, producing patchy patterns and a sparse signal in the images that HOG treats to estimate the $V$ parameter. However, due to images that are sparsely sampled, the HOG algorithm can lose efficiency in assessing the morphological correlation between two identical images. In Fig.~\ref{fig:hogeff}, we show the result of HOG correlating a given image with itself (in purple) while masking out a larger and larger fraction of pixels randomly distributed in the maps. The masked pixels were drawn from Gaussian random realizations with a power-law power spectrum of slope -2.4 to assure a certain spatial correlation among the masked pixels. For comparison, and using the same masked pixels, we also estimated the $V$ parameter from HOG, correlating the previous given map with a completely uncorrelated random map (in gray). For both the purple and gray curves, we computed the mean and standard deviation of $V$ from 100 realizations of the Gaussian random field.

Figure~\ref{fig:hogeff} highlights the fact that the HOG algorithm is not able to distinguish between an intrinsic morphological correlation (purple) and a pure chance correlation (gray), for a 10\% fraction of available pixels in the maps.
Since in the cases of Fields B and C the amount of pixels with a high signal-to-noise ratio in polarization is below 10\%, we claim that the lack of correlation between the LOFAR and EBHIS data in these two fields of view, with a $V$ parameter always compatible with noise, must be related to the employed statistical measure with data at low sensitivity.

\end{document}